\documentclass[a4paper,11pt]{article}

\usepackage{amsfonts,amssymb}
\usepackage{theorem}
\usepackage{epsfig}

\def\G{\Gamma}

\def\hG{{\hat \G}}

\begin{document}

\par
\rightline{IFUM-852-FT}

\vskip 2.0 truecm
\Large
\bf
\centerline{Renormalization of the Non-Linear Sigma Model}
\centerline{in Four Dimensions. A two-loop example.}
\normalsize \rm

\vskip 0.5 truecm
\large
\centerline{Ruggero Ferrari 
\footnote{E-mail address: {\tt ruggero.ferrari@mi.infn.it}} 
and Andrea Quadri \footnote{E-mail address: 
{\tt andrea.quadri@mi.infn.it}}}

\vskip 0.3 truecm
\normalsize
\centerline{Phys. Dept. University of Milan, 
via Celoria 16, 20133 Milan, Italy } 
\centerline{and I.N.F.N., sezione di Milano}

\vskip 0.7  truecm
\normalsize
\bf
\centerline{Abstract}
\rm
\begin{quotation}
\noindent
The renormalization procedure of the non-linear $SU(2)$ sigma model in $D=4$
proposed in Refs.~\cite{Ferrari:2005ii,Ferrari:2005va} 
is here tested in a truly non-trivial
case where the non-linearity of the functional equation is crucial.
The simplest example, where the non-linear term contributes,
is given by the two-loop amplitude 
involving the insertion of two
$\phi_0$ (the constraint of the non-linear sigma model) 
and two flat connections.
In this case we verify the validity of the renormalization procedure:
the recursive subtraction of the pole parts at $D=4$ yields amplitudes
that satisfy the defining functional equation.
As a by-product we give a formal proof that in $D$~dimensions 
(without counterterms) the Feynman rules provide a perturbative
symmetric solution.
\end{quotation}

\newpage

\section{Introduction}

In Refs.~\cite{Ferrari:2005ii,Ferrari:2005va} it was proposed to quantize
the non-linear sigma model in $D=4$ by embedding the pion fields
in a flat connection in order to solve the long-standing problem \cite{Ecker:1972bm}-\cite{Honerkamp:1996va}
of a symmetric removal of the divergences.
The theory is defined by the functional
equation for the connected amplitudes
\begin{eqnarray}
&& 
\!\!\!\!\!\!\!
{\cal S}(W) = 
\Big ( \frac{m_D^2}{2} \partial^\mu \frac{\delta W}{\delta J^\mu_a}
+ g^2 \frac{\delta W}{\delta K_a} K_0 - K_a \frac{\delta W}{\delta K_0} 
- g \epsilon_{abc} K_b \frac{\delta W}{\delta K_c}
 \nonumber \\
&& ~~~~
+ 2 {\cal D}[\frac{\delta W}{\delta J}]^\mu_{ab} J_{b\mu} \Big ) (x) =0 \,
\label{eq.1}
\end{eqnarray}
or for 1-PI amplitudes
\begin{eqnarray}
(\G,\G) & = & 
 \Big ( \frac{m_D^2}{2} \partial^\mu \frac{\delta \G}{\delta J^\mu_a}
+ g^2 \phi_a K_0 + \frac{\delta \G}{\delta K_0} \frac{\delta \G}{\delta \phi_a}
+ g \epsilon_{abc} \frac{\delta \G}{\delta \phi_b} \phi_c \nonumber \\
&& + 2 {\cal D}[\frac{\delta \G}{\delta J}]^\mu_{ab} J_{b\mu} \Big ) (x) =0 \,
\label{eq.2}
\end{eqnarray}
with
\begin{eqnarray}
{\cal D}[X]^\mu_{ab} = \partial^\mu \delta_{ab} - g \epsilon_{abc} X^\mu_c \, .
\label{eq.2.bis}
\end{eqnarray}

Moreover it was conjectured  and shown \cite{Ferrari:2005ii}
in a few examples that
standard perturbation theory in $D$-dimensions gives amplitudes
that satisfy the equation without any subtractions.
The limit to $D=4$ is divergent and needs subtraction of the poles
of the Laurent expansion. This is done by using the properly
normalized 1-PI amplitudes involving only insertions of flat
connections and the constrained field $\phi_0$, i.e. those
amplitudes which are on the top of the hierarchy implicit in
eq.(\ref{eq.2}).

This procedure is consistent if the subtraction procedure 
(use of counterterms in the Feynman rules) does not violate
eq.(\ref{eq.2}) which is a non-linear equation in the bilinear
term
\begin{eqnarray}
\frac{\delta \G}{\delta K_0(x)} \frac{\delta \G}{\delta \phi_a(x)} \, .
\label{eq.4}
\end{eqnarray}

At one-loop level this problem has been addressed in the paper 
\cite{Ferrari:2005va}. 
There it was shown that at one-loop level eq.(\ref{eq.2}) becomes
linear and the counterterms given by the subtraction procedure are a solution
of the equation. 
So at the one-loop level there is no anomaly.
Moreover in the same paper the most general local finite renormalization
compatible with the functional equation 
has been classified (the number of these free
parameters is finite, a property that we indicated as weak power-counting
theorem).

The simplest two-loop example where the bilinear term in eq.(\ref{eq.4})
is non-zero is the four-point-amplitude involving two flat
connections and two constrained fields. In this case
eq.(\ref{eq.2}) becomes
\begin{eqnarray}
&& \frac{m_D^2}{2} \partial^\mu \G^{(2)}_{J^\mu_a J K_0K_0}
+ m_D \G^{(2)}_{\phi_a J K_0K_0} + \G^{(2)}_{K_0K_0K_0} \G^{(0)}_{\phi_a J} 
\nonumber \\
&& + \G^{(1)}_{K_0K_0} \G^{(1)}_{\phi_a K_0 J} = 0 \, .
\label{eq.5}
\end{eqnarray}
In this paper we show that the subtraction procedure for the $D=4$
limit provides amplitudes that satisfy eq.(\ref{eq.5}).
We do not evaluate explicitly the second-order counterterms, rather
we study eq.(\ref{eq.5}) where only the first order counterterms
are introduced and we prove that the local second order
subtraction restores the validity of the equation.

\medskip
The paper is organized as follows. In Sect.~\ref{sec.1} we consider 
the consequences of the introduction of the first order counterterms
on the functional equation at the two-loop level. 
In Sect.~\ref{sec.2} 
we prove that the contributuion of the graphs with no counterterms
cancel exactly. This result is obtained by a quantum action principle
for the unsubtracted amplitudes. In Sect.~\ref{sec.3} we consider the
contribution of all graphs containing one-loop counterterms. 
The result of these two sections allows the evaluation of the breaking
term of the functional equation which turns out to be local.
This is described in Sect.~\ref{sec.4}. We conclude that the breaking term
is removed completely by the subtraction procedure and therefore 
that the functional equation is stable under renormalization in this particular
example.

\section{Breaking of the functional equation}\label{sec.1}

The object of our investigation is the amplitude $\G^{(2)}_{JJK_0K_0}$ at the
two-loop level in the limit $D=4$.
The contribution to this amplitude without insertion of counterterms will be denoted
by $\G^{(2,0)}_{JJK_0K_0}$ while the amplitude with one
counterterms insertion will be denoted by $\G^{(2,1)}_{JJK_0K_0}$. The two
contributions yield an amplitude $\G^{(2,0)}_{JJK_0K_0}+\G^{(2,1)}_{JJK_0K_0}$
which is expected to develop a pole in the limit $D=4$, thus
necessitating a further and last subtraction. 
However before doing this last step we
consider the breaking of the functional equation caused by the 
one-loop counterterm insertion
\begin{eqnarray}
&& 
\frac{m_D^2}{2} \partial^\mu \Big ( \G^{(2,0)}_{J^\mu_a}[JJK_0K_0]+\G^{(2,1)}_{J^\mu_a}[JJK_0K_0] \Big )  
\nonumber \\
&&
+ m_D \Big ( \G^{(2,0)}_{\phi_a}[\phi J K_0K_0] + \G^{(2,1)}_{\phi_a}[\phi J K_0K_0] \Big ) \nonumber \\
&& 
+ \Big ( \G^{(2,0)}_{K_0}[K_0K_0K_0] + \G^{(2,1)}_{K_0}[K_0K_0K_0] \Big ) \G^{(0)}_{\phi_a}[\phi J] \nonumber \\
&&  + \Big ( \G^{(1,0)}_{K_0}[K_0K_0] + \hat \G^{(1)}_{K_0}[K_0K_0] \Big )
     \Big ( \G^{(1,0)}_{\phi_a}[\phi K_0 J] + \hat \G^{(1)}_{\phi_a}[\phi  K_0 J] \Big ) 
\nonumber \\
&& ~~~~~ = \Delta^{(2)}[JJK_0] \, .
\label{sec.1:e1}
\end{eqnarray}

The crucial point is the evaluation of $\Delta^{(2)}$. It is important to show that it is local
and that it is removed by the subtraction procedure. 
It is worth to recall here the subtraction procedure at $D=4$. This is performed
recursively, i.e. after the $(n-1)$-subtraction has been performed the resulting amplitudes
are properly normalized and expanded in a Laurent series at $D=4$.
Finally the pole parts, which are local, are removed by introducing the counterterms
in the effective action. 

It is very crucial to respect the hierarchy and to normalize the amplitudes. The functional equation shows that one has to make finite only the amplitudes involving derivatives w.r.t $J_\mu$ and $K_0$.  In fact all
the other amplitudes are derived by subsequent functional differentiation w.r.t the field $\phi_a$.
For dimensional reasons the amplitudes are normalized by
\begin{eqnarray}
\Big ( \frac{m_D}{m} \Big )^{2(n-1)} \G_{J^{\mu_1}_{a_1 } \dots J^{\mu_n}_{a_n}} =
m^{(n-1)(D-4)} \G_{J^{\mu_1}_{a_1 } \dots J^{\mu_n}_{a_n}} \, .
\label{sec.1:e2}
\end{eqnarray}
Once again we stress that 
the very definition of the theory at $D=4$ crucially depends on the subtraction procedure outlined above
and our goal is to prove
that it does not spoil the functional equation (no anomalies).

\section{Amplitudes without counterterms}\label{sec.2}

It has been conjectured in \cite{Ferrari:2005ii} that the unsubtracted amplitudes
satisfy the functional equation. It is worth then to consider a subset
of the terms appearing in eq.(\ref{sec.1:e1}), i.e. those involving amplitudes
$\G^{(1,0)}$ and $\G^{(2,0)}$. In this section we in fact demonstrate the
correctness of the conjecture by using quantum action principle arguments
\cite{Breitenlohner:1977hr}-\cite{Breitenlohner:1976te}.

We use the generating functional
\begin{eqnarray}
Z[K_a,K_0,J_\mu] = \exp \Big ( {i \left . \G^{(0)}_{int}[\phi,K_0,J] \right |_{\phi_a = \frac{1}{i}\frac{\delta}{\delta K_a}}} \Big ) ~
\exp  \Big ( {\frac{i}{2} \int K_a \Delta_F K_a} \Big ) 
\label{sec.2:e1}
\end{eqnarray}
where tadpole contributions are discharged.
Then we apply the operator ${\cal S}$ in  eq.(\ref{eq.1}) to the connected generating functional 
\begin{eqnarray}
W = -i \ln Z
\label{sec.2:e2}
\end{eqnarray}
and we get
\begin{eqnarray}
{\cal S}(W) & = & i \Big ( \frac{m_D^2}{2} \partial^\mu \frac{\delta \G^{(0)}}{\delta J_\mu^a}
                  + g^2 \phi_a K_0 \nonumber \\
            &   & ~~~ + \frac{\delta \G^{(0)}}{\delta K_0} \frac{\delta \G^{(0)}}{\delta \phi_a} 
                      + g \epsilon_{abc} \frac{\delta \G^{(0)}}{\delta \phi_b} \phi_c \nonumber \\
            &   & ~~~ + 2 {\cal D} [ \frac{\delta \G^{(0)}}{\delta J} ]^\mu_{ab} J_{b\mu} \Big ) \cdot
                  W[K_a,K_0,J_\mu] \, 
\label{sec.2:e3}
\end{eqnarray}
where the $\cdot$ denotes the insertion.
Since
\begin{eqnarray}
(\G^{(0)},\G^{(0)}) & = &  \Big ( \frac{m_D^2}{2} \partial^\mu \frac{\delta \G^{(0)}}{\delta J^\mu_a}
+ g^2 \phi_a K_0 + \frac{\delta \G^{(0)}}{\delta K_0} \frac{\delta \G^{(0)}}{\delta \phi_a}
+ g \epsilon_{abc} \frac{\delta \G^{(0)}}{\delta \phi_b} \phi_c \nonumber \\
&& + 2 {\cal D}[\frac{\delta \G^{(0)}}{\delta J}]^\mu_{ab} J_{b\mu} \Big ) (x) =0
\label{sec.2:e4}
\end{eqnarray}
we obtain
\begin{eqnarray}
{\cal S}(W) = 0 \, .
\label{sec.2:e5}
\end{eqnarray}
All the insertions in the R.H.S. of eq.(\ref{sec.2:e3}) mediated by one-particle states
cancel out. Therefore all the insertions of the composite operators in the R.H.S.
of eq.(\ref{sec.2:e3}) can be replaced by insertions on $\G$. Hence we get
\begin{eqnarray}
&& \!\!\!\!\!\!\!\!\!\!\!\!\!\!\!\!\!\!\!\!\!\!\!\!
i \Big ( \frac{m_D^2}{2} \partial^\mu \frac{\delta \G^{(0)}}{\delta J_\mu^a}
                  + g^2 \phi_a K_0 
                      + \frac{\delta \G^{(0)}}{\delta K_0} \frac{\delta \G^{(0)}}{\delta \phi_a} 
                      + g \epsilon_{abc} \frac{\delta \G^{(0)}}{\delta \phi_b} \phi_c \nonumber \\
&& 
\!\!\!\!\!\!\!\!\!\!\!\!\!\!\!\!\!\!\!\!\!\!\!\!
+ 2 {\cal D} [ \frac{\delta \G^{(0)}}{\delta J} ]^\mu_{ab} J_{b\mu} \Big ) \cdot
                  \G[\phi_a,K_0,J_\mu]  = 0 
\label{sec.2:e6}
\end{eqnarray}
This equation can be used to perform a detailed diagrammatic analysis (see Appendix ~\ref{appA})
of the validity of the functional equation (\ref{eq.2}) at two-loop level for the unsubtracted amplitudes:
\begin{eqnarray}
&& \!\!\!\!\!\!\!\!\!\!\!\!\!\!\!\!\!\!\!\!\!\!\!\!\!
\frac{m_D^2}{2} \partial_\mu \frac{\delta \G^{(2,0)}[JJK_0K_0]}{\delta J^\mu_a}
+ m_D \frac{\delta \G^{(2,0)}[\phi J K_0K_0]}{\delta \phi_a} \nonumber \\
              && 
             \!\!\!\!\!\!\!\!\!\!\!\!\!\!\!\!\!
              + \frac{\delta \G^{(2,0)}[K_0K_0K_0]}{\delta K_0} 
                      \frac{\delta \G^{(0)}[\phi J]}{\delta \phi_a}
                    + \frac{\delta  \G^{(1,0)}[K_0K_0]}{\delta K_0} \frac{\delta \G^{(1,0)}[\phi_a J K_0]}{\delta \phi_a} = 0 \, .
\label{sec.2:e7}
\end{eqnarray}

\section{Amplitudes with one-loop counterterms}\label{sec.3}

The surprising result of this section is that the sum of all the
amplitudes carrying a one-loop counterterm and that can potentially
give a non-local contribution to $\Delta^{(2)}$ in eq.(\ref{sec.1:e1})
do in fact sum up to zero:
\begin{eqnarray}
&& 
\frac{m_D^2}{2} \partial^\mu \G^{(2,1)}_{J^\mu_a}[JJ K_0 K_0] + m_D \G^{(2,1)}_{\phi_a}[K_0 K_0 J \phi]
\nonumber \\
&& 
+ \G^{(2,1)}_{K_0} [K_0 K_0 K_0] \G^{(0)}_{\phi_a} [\phi J]
 + \G^{(1,0)}_{K_0}[K_0 K_0] \hat \G^{(1)}_{\phi_a}[\phi K_0 J] \nonumber \\
&& + \hG^{(1)}_{K_0} [K_0K_0] \G^{(1,0)}_{\phi_a} [J K_0 \phi] = 0 \, .
\label{sec.3:e1}
\end{eqnarray}
We first give a general argument based on the quantum action principle for the
generating functional of the amplitudes which now includes the counterterms at one-loop among the
Feynman rules. Then eq.(\ref{sec.2:e1}) now becomes
\begin{eqnarray}
\!\!\!\!\!\!
Z_{1R}[K_a, K_0, J] = \exp \Big ( \left . ( i \G^{(0)}_{int} + i \hat \G^{(1)} ) \right |_{\phi_a = \frac{1}{i}\frac{\delta}{\delta K_a}}  \Big )  ~
\exp  \Big ( {\frac{i}{2} \int K_a \Delta_F K_a} \Big ) \, .
\label{sec.3:e2}
\end{eqnarray}
The functional equation for the connected amplitudes $W_{1R} = -i \ln Z_{1R}$ now becomes
\begin{eqnarray}
{\cal S}(W_{1R}) & = & \Big ( \frac{m_D^2}{2} \partial^\mu \frac{\delta [ \G^{(0)} + \hat \G^{(1)} ]}{\delta J^\mu_a(x)}
+ g^2 \phi_a(x) K_0(x) \nonumber \\
                 &   & \!\!\!\!\!\!\!\!\!\!\!\! + \frac{\delta [ \G^{(0)} + \hat \G^{(1)} ]}{\delta K_0(x)} 
                         \frac{\delta [ \G^{(0)} + \hat \G^{(1)} ]}{\delta \phi_a(x)}
+ 2 {\cal D}\Big [ \frac{\delta [ \G^{(0)} + \hat \G^{(1)} ]}{\delta J} \Big ]_\mu^{ab} J^\mu_b(x) \Big )
                        \cdot W_{1R} \, . \nonumber \\
\label{sec.3:e3}
\end{eqnarray}

The counterterms obey the linearized form of eq.(\ref{eq.2}) 
(see Refs.~\cite{Ferrari:2005ii,Ferrari:2005va}):
\begin{eqnarray}
\!\!\!\!\!\!\!\!\!\!\!\!\!
  \frac{m_D^2}{2} \partial^\mu \frac{\delta \hat \G^{(1)}}{\delta J^\mu_a(x)} 
+ \frac{\delta \G^{(0)}}{\delta K_0(x)} \frac{\delta \hat \G^{(1)}}{\delta \phi_a(x)}
+ \frac{\delta \hat \G^{(1)}}{\delta K_0(x)} \frac{\delta \G^{(0)}}{\delta \phi_a(x)}
+ 2 {\cal D} \Big [ \frac{\delta \hat \G^{(1)}}{\delta J} \Big ]_\mu^{ab} J^\mu_b(x)  =0 \, .
\label{sec.3:e4}
\end{eqnarray}
Therefore as a consequence of the quantum action principle in eq.(\ref{sec.3:e3})
and of eqs.(\ref{sec.2:e4}) and (\ref{sec.3:e4}) 
we get
\begin{eqnarray}
{\cal S}(W_{1R}) = \frac{\delta \hat \G^{(1)}}{\delta K_0(x)} \frac{\delta \hat \G^{(1)}}{\delta \phi_a(x)} \cdot  W_{1R} \, .
\label{sec.3:e5}
\end{eqnarray} 

This result can be explicitly verified in our particular example. In appendix ~\ref{appB}
we prove diagrammatically the validity of eq.(\ref{sec.3:e1}) and consequently that of eq.(\ref{sec.3:e5})
by the evaluation of $\Delta^{(2)}$ in eq.(\ref{sec.1:e1}).
Two different checks are needed. 
First one verifies that 
indeed the insertion of the L.H.S. in eq.(\ref{sec.3:e4})  is zero:
\begin{eqnarray}
&& \!\!\!\!\!\!\!\!\!\!\!\!\!
\Big ( \frac{m_D^2}{2} \partial^\mu \frac{\delta \hat \G^{(1)}}{\delta J^\mu_a(x)} 
+ \frac{\delta \G^{(0)}}{\delta K_0(x)} \frac{\delta \hat \G^{(1)}}{\delta \phi_a(x)}
+ \frac{\delta \hat \G^{(1)}}{\delta K_0(x)} \frac{\delta \G^{(0)}}{\delta \phi_a(x)} 
+ 2 {\cal D} \Big [ \frac{\delta \hat \G^{(1)}}{\delta J} \Big ]_\mu^{ab} J^\mu_b(x) \Big )  \nonumber \\
&& ~~~~ \cdot W_{1R} = 0 \, .
\label{sec.3:e6}
\end{eqnarray}
Again all the insertions in the L.H.S. of eq.(\ref{sec.3:e6}) mediated by one-particle states cancel out.
Therefore all the insertions of the composite operators in the R.H.S. of eq.(\ref{sec.3:e6})
can be replaced by insertions on $\G$. Hence we get
\begin{eqnarray}
&& \!\!\!\!\!\!\!\!\!\!\!\!\!
\Big ( \frac{m_D^2}{2} \partial^\mu \frac{\delta \hat \G^{(1)}}{\delta J^\mu_a(x)} 
+ \frac{\delta \G^{(0)}}{\delta K_0(x)} \frac{\delta \hat \G^{(1)}}{\delta \phi_a(x)}
+ \frac{\delta \hat \G^{(1)}}{\delta K_0(x)} \frac{\delta \G^{(0)}}{\delta \phi_a(x)} 
+ 2 {\cal D} \Big [ \frac{\delta \hat \G^{(1)}}{\delta J} \Big ]_\mu^{ab} J^\mu_b(x) \Big )  \nonumber \\
&& ~~~~ \cdot \G_{1R} = 0 \, .
\label{sec.3:e7}
\end{eqnarray}
Moreover one  needs to check that the insertion of the L.H.S. of eq.(\ref{sec.2:e4}) 
is  zero also at one-loop level, i.e. 
\begin{eqnarray}
(\G^{(0)},\G^{(0)}) \cdot W_{1R} = 0 \, .
\label{sec.3:e8,bis}
\end{eqnarray}
In the above equation  we can again restrict ourselves to the 1-PI amplitudes since the insertions mediated
by one-particles states in the L.H.S. cancel out, so that
\begin{eqnarray}
(\G^{(0)},\G^{(0)}) \cdot \G_{1R} = 0 \, .
\label{sec.3:e8}
\end{eqnarray}

The diagrammatic evaluation of the L.H.S. eq.(\ref{sec.3:e1}) is finally achieved by 
a combined use of eq.(\ref{sec.3:e7}) and eq.(\ref{sec.3:e8}).

\section{Removal of the breaking $\Delta^{(2)}[JK_0K_0]$}\label{sec.4}

The result of the previous sections has shown that the breaking of the functional equation
at the two-loop level after the introduction of the one-loop level counterterms is given by
\begin{eqnarray}
\Delta^{(2)} = \frac{\delta \hat \G^{(1)}}{\delta K_0(x)} \frac{\delta \hat \G^{(1)}}{\delta \phi_a(x)} \cdot \G_{1R} \, .
\label{sec.4:e1}
\end{eqnarray}
In order to proceed further we specialize to the case we are dealing with
\begin{eqnarray}
&& 
\!\!\!\!\!\!\!\!\!\!\!\!\!\!\!\!\!\!\!\!\!\!\!\!\!\!\!\!
\Delta^{(2)}[J K_0 K_0] =  \hat \G^{(1)}_{K_0(x)}[K_0K_0] \hat \G^{(1)}_{\phi_a(x)}[\phi J K_0] \cdot 1 \nonumber \\
&& \!\!\!\!\!\!\!\!\!\!\!\!\!\!\!\!\!\!\!\! = \Big ( \frac{1}{D-4} \Big )^2 \frac{6 g^8}{(4 \pi)^4} \frac{1}{m^4 m_D^2} 
                              K_0(x) ( -4 \partial^\mu K_0(x) J_{a\mu}(x) - K_0(x) \partial J_a(x) ) \, .
\label{sec.4:e2}
\end{eqnarray}
This is of course a local insertion and then the last very crucial point is to verify that after the
subtraction procedure it disappears (not even finite parts are left over).
In order to prove that this is indeed the case one must remember that the subtraction of the poles
has to be performed on the normalized amplitudes as stated in eq.(\ref{sec.1:e2}).
Thus we have to expand in a Laurent series both sides of the equation
\begin{eqnarray}
&& \!\!\!\!\!\!\!\!\! 
\frac{1}{m^2} \Big ( \frac{m_D}{m} \Big )^2 \Big ( \frac{m_D^2}{2} \partial^\mu \frac{\delta \G^{(2)}[JJK_0K_0]}{\delta J^\mu_a(x)} 
              + m_D \frac{\delta \G^{(2)}[\phi J K_0 K_0]}{\delta \phi_a(x)} \nonumber \\
&& ~~~~ + \frac{\delta \G^{(1)}[K_0 K_0]}{\delta K_0(x)} \frac{\delta \G^{(1)}[\phi K_0 J]}{\delta \phi_a(x)} 
        + \frac{\delta \G^{(2)}[K_0 K_0 K_0]}{\delta K_0(x)} \frac{\delta \G^{(0)}[\phi J]}{\delta \phi_a(x)} \Big ) 
\nonumber \\
&& = \frac{1}{m^2} \Big ( \frac{m_D}{m} \Big )^2 \Delta^{(2)}[K_0 K_0 J] \, .              
\label{sec.4:e3}
\end{eqnarray}

In the R.H.S. $m_D^2$ cancels out and therefore the R.H.S. contains a pure double pole with no finite
parts left over. The pole part disappears after the subtraction procedure has been performed on the L.H.S.
of the above equation.
Once again we stress that the subtraction procedure has to be applied in the L.H.S. of
eq.(\ref{sec.4:e3}) to each of the terms $\Big ( \frac{m_D}{m} \Big )^4 \G^{(2)}[JJK_0K_0]$, 
$\Big ( \frac{m_D}{m} \Big )^3 \G^{(2)}[\phi J K_0 K_0]$
and $\Big ( \frac{m_D}{m} \Big ) \G^{(2)}[K_0 K_0 K_0]$ (notice that $\frac{\delta \G^{(0)}[\phi J]}{\delta \phi_a(x)} = \frac{2}{m_D}
J^\mu_a \partial_\mu \phi_a$).

\section{Conclusions and outlook}

The aim of this paper is to show on a specific example 
that the subtraction procedure at $D=4$
is symmetric, i.e. the functional equation is stable under renormalization at the two-loop level.
The proof is based on the evaluation of the breaking term at two loops after the introduction
of first-order counterterms. We have been able to show that the breaking is a local insertion.
Moreover it is subtracted completely when we remove the overall poles in $D=4$ from the two-loop amplitudes
corrected by the insertion of one-loop counterterms.

The technique used can be probably applied to a general case since we use a recursive method. As a by-product
we have also shown that the perturbative series in generic dimension $D$ satisfies the functional equation.
After these results the theory looks promising and one can think to some phenomenological applications
and to a more general approach (systematic use of cohomological methods to classify finite renormalizations)
to the renormalization of the non-linear sigma model.

Many open questions can be addressed at this point. We consider of particular interest the possibility
to interpret finite renormalizations as a kind of deformation of the geometry of the $\phi$-manifold
\cite{Brezin:1976ap}-\cite{Alvarez-Gaume:1981hn}
also in $D=4$.

\appendix
\section{Diagrammatic analysis of the functional equation for unsubtracted amplitudes}\label{appA}

We evaluate the insertion in the L.H.S. of eq.(\ref{sec.2:e6}) in the relevant 1-PI sector spanned 
by two $K_0$ and one $J_\mu$ (see eq.(\ref{sec.2:e7})).

The functional equation (\ref{sec.2:e4}) yields
\begin{eqnarray}
\frac{m_D^2}{2} \partial^\mu \frac{\delta \G^{(0)}[J\phi\phi\phi]}{\delta J^\mu_a(x)}
+ m_D \frac{\delta \G^{(0)}[\phi\phi\phi\phi]}{\delta \phi_a(x)} 
+ \frac{\delta \G^{(0)}[K_0\phi\phi]}{\delta K_0(x)}
  \frac{\delta \G^{(0)}[\phi\phi]}{\delta \phi_a(x)}
= 0 \, ,
\label{appA:1}
\end{eqnarray}
\begin{eqnarray}
\frac{m_D^2}{2} \partial^\mu \frac{\delta \G^{(0)}[J\phi\phi]}{\delta J^\mu_a(x)}
+ g \epsilon_{abc} \frac{\delta \G^{(0)}[\phi\phi]}{\delta \phi_b(x)} \phi_c(x) = 0 \, ,
\label{appA:2}
\end{eqnarray}
\begin{eqnarray}
m_D \frac{\delta \G^{(0)}[K_0\phi\phi\phi\phi]}{\delta \phi_a(x)}
+ \frac{\delta \G^{(0)}[K_0\phi\phi]}{\delta K_0(x)} 
  \frac{\delta \G^{(0)}[K_0\phi\phi]}{\delta \phi_a(x)} = 0 \, ,
\label{appA:3}
\end{eqnarray}
\begin{eqnarray}
&& 
\!\!\!\!\!\!\!\!\!\!\!\!\!\!\!\!\!\!\!\!\!
m_D \frac{\delta \G^{(0)}[J\phi\phi\phi\phi\phi]}{\delta \phi_a(x)} 
+ \frac{\delta \G^{(0)}[K_0\phi\phi]}{\delta K_0(x)} \frac{\delta \G^{(0)}[J\phi\phi\phi]}{\delta \phi_a(x)} 
\nonumber \\
&& + \frac{\delta \G^{(0)}[K_0\phi\phi\phi\phi]}{\delta K_0(x)} \frac{\delta \G^{(0)}[J\phi]}{\delta \phi_a(x)} = 0 \, , 
\label{appA:4}
\end{eqnarray}
\begin{eqnarray}
&& m_D \frac{\delta \G^{(0)}[J\phi\phi\phi]}{\delta \phi_a(x)} 
+ \frac{\delta \G^{(0)}[K_0\phi\phi]}{\delta K_0(x)} \frac{\delta \G^{(0)}[J\phi]}{\delta \phi_a(x)}
\nonumber \\
&& ~~~~~~~ + g \epsilon_{abc} \frac{\delta \G^{(0)}[J\phi\phi]}{\delta \phi_b(x)} \phi_c(x) 
-2 g \epsilon_{abc} \frac{\delta \G^{(0)}[J\phi\phi]}{\delta J^\mu_c(x)} J^\mu_b(x) = 0 \, .
\label{appA:5}
\end{eqnarray}

\begin{figure}
\begin{center}
\epsfig{file=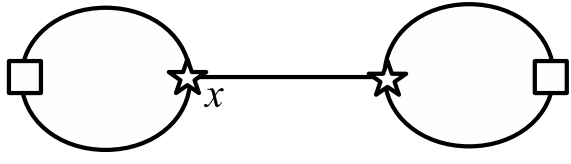,width=8cm}
\caption{Spurious (non 1-PI) contribution in the insertion of eq.(\ref{appA:ins.1}) (stars and boxes are insertions of flat connections and of $\phi_0$ respectively).}
\label{fig.1}
\end{center}
\end{figure}

From eq.(\ref{appA:1}) we obtain (in $\langle \cdot \rangle$ 
we omit the $T$-symbol in order to simplify
the notations)
\begin{eqnarray}
&& \!\!\!\!\! 
\frac{1}{2} \Big \langle \Big ( \frac{m_D^2}{2} \partial^\mu \frac{\delta \G^{(0)}[J\phi\phi\phi]}{\delta J^\mu_a(x)}
+ m_D \frac{\delta \G^{(0)}[\phi\phi\phi\phi]}{\delta \phi_a(x)} 
+ \frac{\delta \G^{(0)}[K_0\phi\phi]}{\delta K_0(x)}
  \frac{\delta \G^{(0)}[\phi\phi]}{\delta \phi_a(x)} \Big ) \nonumber \\
&& ~~~~ \G^{(0)}[J\phi\phi\phi] \G^{(0)}[K_0\phi\phi]
  \G^{(0)}[K_0\phi\phi] \Big \rangle \nonumber \\
&& \!\!\!\!\! -  \Big \langle \Big \langle 
    \Big ( \frac{m_D^2}{2} \partial^\mu \frac{\delta \G^{(0)}[J\phi\phi\phi]}{\delta J^\mu_a(x)}
+ m_D \frac{\delta \G^{(0)}[\phi\phi\phi\phi]}{\delta \phi_a(x)} 
+ \frac{\delta \G^{(0)}[K_0\phi\phi]}{\delta K_0(x)}
  \frac{\delta \G^{(0)}[\phi\phi]}{\delta \phi_a(x)} \Big ) \G^{(0)}[K_0\phi\phi] \Big \rangle \nonumber \\
&& ~~~~~~~ \Big \langle \G^{(0)}[J\phi\phi\phi] \G^{(0)}[K_0\phi\phi] \Big \rangle \Big \rangle = 0 \, .
\label{appA:ins.1} 
\end{eqnarray}
The second term in the above equation subtracts the connected but not 1-PI
contribution, depicted in Fig.~\ref{fig.1}, which enters in the set
of contractions in the first term of eq.(\ref{appA:ins.1}).

Furthermore we obtain from eq.(\ref{appA:2}) 
\begin{eqnarray}
&& \Big \langle \Big ( \frac{m_D^2}{2} \partial^\mu \frac{\delta \G^{(0)}[J\phi\phi]}{\delta J^\mu_a(x)}
+ g \epsilon_{abc} \frac{\delta \G^{(0)}[\phi\phi]}{\delta \phi_b(x)} \phi_c(x) \Big ) \nonumber \\
&&  ~~~~~~~~~~~~~~  \G^{(0)}[J\phi\phi] 
\G^{(0)}[K_0\phi\phi\phi\phi] \G^{(0)}[K_0\phi\phi] \Big \rangle = 0 \, 
\nonumber \\
\label{appA:ins.2} 
\end{eqnarray}
and also
\begin{eqnarray}
&& \frac{i}{2} \Big \langle \Big ( \frac{m_D^2}{2} \partial^\mu \frac{\delta \G^{(0)}[J\phi\phi]}{\delta J^\mu_a(x)}
+ g \epsilon_{abc} \frac{\delta \G^{(0)}[\phi\phi]}{\delta \phi_b(x)} \phi_c(x) \Big ) \nonumber \\
&& ~~~~~~~~~~~~~~ \G^{(0)}[J\phi\phi] \G^{(0)}[K_0\phi\phi] \G^{(0)}[K_0\phi\phi] 
              \G^{(0)}[\phi\phi\phi\phi] \Big \rangle = 0 \, .
\label{appA:ins.3}
\end{eqnarray}

From eq.(\ref{appA:3}) we get
\begin{eqnarray}
&& 
\!\!\!\!\!\!
-i \Big \langle \Big ( m_D \frac{\delta \G^{(0)}[K_0\phi\phi\phi\phi]}{\delta \phi_a(x)} 
+ \frac{\delta \G^{(0)}[K_0\phi\phi]}{\delta K_0(x)} 
  \frac{\delta \G^{(0)}[K_0\phi\phi]}{\delta \phi_a(x)} \Big ) 
\G^{(0)}[K_0\phi\phi] \G^{(0)}[J\phi\phi\phi] \Big \rangle  = 0 \, . \nonumber \\
\label{appA:ins.4}
\end{eqnarray}

From eq.(\ref{appA:4}) we get
\begin{eqnarray}
&& 
\!\!\!\!\!\!
-\frac{i}{2} \Big \langle \Big ( m_D \frac{\delta \G^{(0)}[J\phi\phi\phi\phi\phi]}{\delta \phi_a(x)} 
+ \frac{\delta \G^{(0)}[K_0\phi\phi]}{\delta K_0(x)} \frac{\delta \G^{(0)}[J\phi\phi\phi]}{\delta \phi_a(x)} + \frac{\delta \G^{(0)}[K_0\phi\phi\phi\phi]}{\delta K_0(x)} \frac{\delta \G^{(0)}[J\phi]}{\delta \phi_a(x)}
\Big ) \nonumber \\ 
&& ~~~~ \G^{(0)}[K_0\phi\phi] \G^{(0)}[K_0\phi\phi] \Big \rangle = 0 \, .
\label{appA:ins.5}
\end{eqnarray}

In order to decompose further the diagrams contributing to the above insertions we need
the fact that 
\begin{eqnarray}
\square_x ~ \Delta_F(x-y) = -i \delta^D(x-y) \, . 
\label{appA:simp.1}
\end{eqnarray}
By using eq.(\ref{appA:simp.1}) the third term in the first line of eq.(\ref{appA:ins.1}) becomes
\begin{eqnarray}
&& \!\!\!\!\!\! \frac{1}{2} \Big \langle \frac{\delta \G^{(0)}[K_0\phi\phi]}{\delta K_0(x)}
  \frac{\delta \G^{(0)}[\phi\phi]}{\delta \phi_a(x)} \G^{(0)}[J\phi\phi\phi] \G^{(0)}[K_0\phi\phi]
  \G^{(0)}[K_0\phi\phi] \Big \rangle  \nonumber \\
&&  = -\frac{1}{2} \Big \langle \frac{\delta \G^{(0)}[K_0\phi\phi]}{\delta K_0(x)}
   \square \phi_a(x) \G^{(0)}[J\phi\phi\phi] \G^{(0)}[K_0\phi\phi]
  \G^{(0)}[K_0\phi\phi] \Big \rangle \nonumber \\
&& =   i \Big \langle 
                  \frac{\delta \G^{(0)}[K_0\phi\phi]}{\delta \phi_a(x)}
                  \G^{(0)}[K_0\phi\phi]   
              \frac{\delta \G^{(0)}[K_0\phi\phi]}{\delta K_0(x)}
              \G^{(0)}[J\phi\phi\phi]  \Big \rangle  \nonumber \\
&& ~~ + \frac{1}{2} i \Big \langle \G^{(0)}[K_0\phi\phi] \G^{(0)}[K_0\phi\phi]   
              \frac{\delta \G^{(0)}[K_0\phi\phi]}{\delta K_0(x)} 
              \frac{\delta \G^{(0)}[J\phi\phi\phi]}{\delta \phi_a(x)}  
              \Big \rangle \, . 
\label{appA:simp.2}
\end{eqnarray}
In a similar fashion the third term in the third line of the L.H.S. of eq.(\ref{appA:ins.1}) can be simplified
as follows:
\begin{eqnarray}
&& \!\!\!\!\!\!\!\!\!\! -  \Big \langle \Big \langle \frac{\delta \G^{(0)}[K_0\phi\phi]}{\delta K_0(x)}
  \frac{\delta \G^{(0)}[\phi\phi]}{\delta \phi_a(x)} 
  \G^{(0)}[K_0\phi\phi] \Big \rangle 
  \Big \langle \G^{(0)}[J\phi\phi\phi] \G^{(0)}[K_0\phi\phi] \Big \rangle \Big \rangle \nonumber \\
&& =  \Big \langle \Big \langle \frac{\delta \G^{(0)}[K_0\phi\phi]}{\delta K_0(x)}
  \square \phi_a(x)
  \G^{(0)}[K_0\phi\phi] \Big \rangle 
  \Big \langle \G^{(0)}[J\phi\phi\phi] \G^{(0)}[K_0\phi\phi] \Big \rangle \Big \rangle \nonumber \\
&& =  - i \Big \langle \Big \langle   \frac{\delta \G^{(0)}[K_0\phi\phi]}{\delta K_0(x)}    																	      \G^{(0)}[K_0\phi\phi] \Big \rangle \Big \langle \G^{(0)}[J\phi\phi\phi] \frac{\delta \G^{(0)}[K_0\phi\phi]}{\delta \phi_a(x)} \Big \rangle \Big \rangle 
\label{appA:n1}
\end{eqnarray}
where all contributions including tadpoles have been discarded.

The second term in the L.H.S. of eq.(\ref{appA:ins.2}) becomes
\begin{eqnarray}
&& \!\!\!\!\!\! 
 \Big \langle g \epsilon_{abc} \frac{\delta \G^{(0)}[\phi\phi]}{\delta \phi_b(x)} \phi_c(x)   \G^{(0)}[J\phi\phi] \G^{(0)}[K_0\phi\phi\phi\phi] \G^{(0)}[K_0\phi\phi] \Big \rangle \nonumber \\
&& = - \Big \langle g \epsilon_{abc} \square \phi_b(x)\phi_c(x)   \G^{(0)}[J\phi\phi] \G^{(0)}[K_0\phi\phi\phi\phi] \G^{(0)}[K_0\phi\phi] \Big \rangle  \nonumber \\
&& =  i \Big \langle g \epsilon_{abc} \frac{\delta \G^{(0)}[J\phi\phi]}{\delta \phi_b(x)} \phi_c(x)
\G^{(0)}[K_0\phi\phi\phi\phi] \G^{(0)}[K_0\phi\phi] \Big \rangle 
\label{appA:simp.3}
\end{eqnarray}
since by the functional equation 
\begin{eqnarray}
&& g \epsilon_{abc} \frac{\delta \G^{(0)}[K_0\phi\phi\phi\phi]}{\delta \phi_b(x)} \phi_c(x) = 0 \, , \nonumber \\
&& g \epsilon_{abc} \frac{\delta \G^{(0)}[K_0\phi\phi]}{\delta \phi_b(x)} \phi_c(x) = 0 \, .
\label{appA:simp.3.1}
\end{eqnarray}
The second term in the L.H.S. of eq.(\ref{appA:ins.3}) becomes
\begin{eqnarray}
&& \!\!\!\!\!\! \frac{i}{2} \Big \langle  g \epsilon_{abc} \frac{\delta \G^{(0)}[\phi\phi]}{\delta \phi_b(x)} \phi_c(x)  \G^{(0)}[J\phi\phi] \G^{(0)}[K_0\phi\phi] \G^{(0)}[K_0\phi\phi] \G^{(0)}[\phi\phi\phi\phi] \Big \rangle  
\nonumber \\
&& = - \frac{i}{2} \Big \langle  g \epsilon_{abc} \square \phi_b(x) \phi_c(x)  \G^{(0)}[J\phi\phi] \G^{(0)}[K_0\phi\phi] \G^{(0)}[K_0\phi\phi] \G^{(0)}[\phi\phi\phi\phi] \Big \rangle  
\nonumber \\
&& =  - \frac{1}{2} \Big \langle g \epsilon_{abc} \frac{\delta \G^{(0)}[J\phi\phi]}{\delta \phi_b(x)} \phi_c(x)
\G^{(0)}[K_0\phi\phi] \G^{(0)}[K_0\phi\phi] \G^{(0)}[\phi\phi\phi\phi] \Big \rangle 
\label{appA:simp.4}
\end{eqnarray}
by the second of eqs.(\ref{appA:simp.3.1}) and the equation
\begin{eqnarray}
&& g \epsilon_{abc} \frac{\delta \G^{(0)}[\phi\phi\phi\phi]}{\delta \phi_b(x)} \phi_c(x) = 0 \, ,
\label{appA:simp.4.1}
\end{eqnarray}
which also follows from the functional equation for $\G^{(0)}$.

It is also convenient to replace  the insertion of 
the composite operator $g \epsilon_{abc} \frac{\delta \G^{(0)}[J\phi\phi]}{\delta \phi_b(x)} \phi_c(x)$
in eqs.(\ref{appA:simp.3}) and (\ref{appA:simp.4})
with that of 
$$- m_D \frac{\delta \G^{(0)}[J\phi\phi\phi]}{\delta \phi_a(x)} 
  - \frac{\delta \G^{(0)}[K_0\phi\phi]}{\delta K_0(x)} \frac{\delta \G^{(0)}[J\phi]}{\delta \phi_a(x)}
  + 2 g \epsilon_{abc} \frac{\delta \G^{(0)}[J\phi\phi]}{\delta J^\mu_c(x)} J^\mu_b(x) $$
by using eq.(\ref{appA:5}). 
We now notice that the relevant unsubtracted amplitudes 
give rise to the following contractions:
\begin{eqnarray}
\!\!\!\!\!\!\!\!
i 
\frac{m_D^2}{2} \partial^\mu \frac{\delta}{\delta J^\mu_a(x)}
\G^{(2,0)}[J J K_0 K_0] & = & \frac{1}{2} \Big \langle 
\frac{m_D^2}{2} \partial^\mu \frac{\delta \G^{(0)}[J\phi\phi\phi] }{\delta J^\mu_a(x)} 
 \G^{(0)}[J\phi\phi\phi] \G^{(0)}[K_0\phi\phi] \G^{(0)}[K_0\phi\phi]  
\Big \rangle \nonumber \\
                              &   & 
\!\!\!\!\!\!\!\!\!\!\!\!\!\!\!\!\!\!\!\!\!\!\!\!\!\!\!\!\!\!\!\!\!\!\!\!
+ \Big \langle \frac{m_D^2}{2} \partial^\mu \frac{\delta \G^{(0)}[J\phi\phi]}{\delta J^\mu_a(x)} 
\G^{(0)}[J\phi\phi] \G^{(0)}[K_0\phi\phi\phi\phi] \G^{(0)}[K_0\phi\phi] \Big \rangle \nonumber \\
                              &   & 
\!\!\!\!\!\!\!\!\!\!\!\!\!\!\!\!\!\!\!\!\!\!\!\!\!\!\!\!\!\!\!\!\!\!\!\!
+\frac{i}{2} \Big \langle 
\frac{m_D^2}{2} \partial^\mu \frac{\delta \G^{(0)}[J\phi\phi]}{\delta J^\mu_a(x)}
 \G^{(0)}[J\phi\phi] \G^{(0)}[K_0\phi\phi] \G^{(0)}[K_0\phi\phi] \G^{(0)}[\phi\phi\phi\phi] \Big \rangle \nonumber \\
                              &   & 
\!\!\!\!\!\!\!\!\!\!\!\!\!\!\!\!\!\!\!\!\!\!\!\!\!\!\!\!\!\!\!\!\!\!\!\!
- \Big \langle \Big \langle \frac{m_D^2}{2} \partial^\mu \frac{\delta \G^{(0)}[J\phi\phi\phi]  }{\delta J^\mu_a(x)}\G^{(0)}[K_0\phi\phi] \Big \rangle
                                                    \Big \langle \G^{(0)}[J\phi\phi\phi] \G^{(0)}[K_0\phi\phi] \Big \rangle 
                                            \Big \rangle \, , \nonumber \\
\label{appA:amp.1}
\end{eqnarray}
\begin{eqnarray}
i m_D\frac{\delta}{\delta \phi_a(x)}
 \G^{(2,0)}[\phi J K_0 K_0] & = & \frac{1}{2} \Big \langle m_D\frac{\delta \G^{(0)}[\phi\phi\phi\phi]}{\delta \phi_a(x)}  \G^{(0)}[K_0\phi\phi] \G^{(0)}[K_0\phi\phi] \G^{(0)}[J\phi\phi\phi] \Big \rangle \nonumber \\
&   & + \frac{1}{2} \Big \langle \G^{(0)}[\phi\phi\phi\phi] \G^{(0)}[K_0\phi\phi] \G^{(0)}[K_0\phi\phi] ~ m_D\frac{\delta \G^{(0)}[J\phi\phi\phi]}{\delta \phi_a(x)}  \Big \rangle \nonumber \\
&   & -i \Big \langle  m_D\frac{\delta \G^{(0)}[K_0\phi\phi\phi\phi]}{\delta \phi_a(x)}  \G^{(0)}[K_0\phi\phi] 
\G^{(0)}[J\phi\phi\phi] \Big \rangle \nonumber \\
&   & -i \Big \langle \G^{(0)}[K_0\phi\phi\phi\phi]  \G^{(0)}[K_0\phi\phi] 
~ m_D\frac{\delta  \G^{(0)}[J\phi\phi\phi] }{\delta \phi_a(x)}\Big \rangle \nonumber \\
&   & -\frac{i}{2} \Big \langle m_D 
\frac{\delta \G^{(0)}[J \phi\phi\phi\phi\phi]}{\delta \phi_a(x)} \G^{(0)}[K_0\phi\phi] \G^{(0)}[K_0\phi\phi] \Big \rangle \nonumber \\
&   & -  \Big \langle \Big \langle m_D \frac{\delta \G^{(0)}[J\phi\phi\phi]}{\delta \phi_a(x)} \G^{(0)}[K_0\phi\phi] \Big \rangle 
             \Big \langle \G^{(0)}[\phi\phi\phi\phi] \G^{(0)}[K_0\phi\phi] \Big \rangle \Big \rangle  
                  \nonumber \\
&   & -  \Big \langle \Big \langle \G^{(0)}[J\phi\phi\phi] \G^{(0)}[K_0\phi\phi] \Big \rangle 
             \Big \langle 
m_D \frac{\delta \G^{(0)}[\phi\phi\phi\phi]}{\delta \phi_a(x)} \G^{(0)}[K_0\phi\phi] \Big \rangle \Big \rangle  
                  \nonumber \\
\label{appA:amp.2}
\end{eqnarray}   
and
\begin{eqnarray}
&& i \frac{\delta \G^{(1,0)}[K_0K_0]}{\delta K_0(x)} 
\frac{\delta \G^{(1,0)}[\phi K_0 J]}{\delta \phi_a(x)} 
 = \nonumber \\
&& ~~~~~~~~~~  -i  \Big \langle \frac{\delta \G^{(0)}[K_0\phi\phi]}{\delta K_0(x)} \G^{(0)}[K_0\phi\phi] \Big \rangle \Big \langle \frac{\delta \G^{(0)}[J\phi\phi\phi]}{\delta \phi_a(x)} \G^{(0)}[K_0\phi\phi] \Big \rangle \, , ~~~~
 \nonumber \\
\label{appA:amp.3}
\end{eqnarray}
\begin{eqnarray}
&&
i \frac{\delta \G^{(2,0)}[K_0 K_0 K_0]}{\delta K_0(x)} \frac{\delta \G^{(0)}[\phi J]}{\delta \phi_a(x)} 
 = \nonumber \\
&& ~~~  
\frac{1}{2} \Big \langle \G^{(0)}[\phi\phi\phi\phi] \frac{\delta \G^{(0)}[K_0\phi\phi]}{\delta K_0(x)} \G^{(0)}[K_0\phi\phi] \G^{(0)}[K_0\phi\phi] \Big \rangle \frac{\delta \G^{(0)}[\phi J]}{\delta \phi_a(x)}  
\nonumber \\
&& ~~~ - \frac{i}{2} \Big \langle \frac{\delta \G^{(0)}[K_0\phi\phi\phi\phi]}{\delta K_0(x)} \G^{(0)}[K_0\phi\phi] 
\G^{(0)}[K_0\phi\phi] \Big \rangle \frac{\delta \G^{(0)}[\phi J]}{\delta \phi_a(x)}  
\nonumber \\
&& ~~~  - i \Big \langle \G^{(0)}[K_0\phi\phi\phi\phi] \frac{\delta \G^{(0)}[K_0 \phi\phi]}{\delta K_0(x)} 
\G^{(0)}[K_0\phi\phi] \Big \rangle \frac{\delta \G^{(0)}[\phi J]}{\delta \phi_a(x)}  \, .
\label{appA:amp.4}
\end{eqnarray}

\begin{figure}
\begin{center}
\epsfig{file=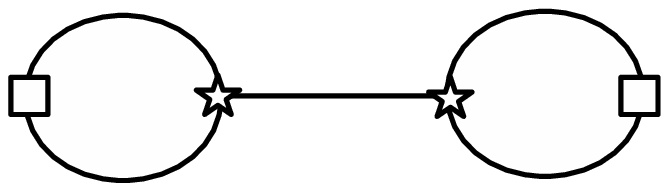,width=6cm}
\caption{Spurious (non 1-PI) contribution to $\G^{(2,0)}[J JK_0K_0]$.}
\label{fig.2}
\end{center}
\end{figure}

\begin{figure}
\begin{center}
\epsfig{file=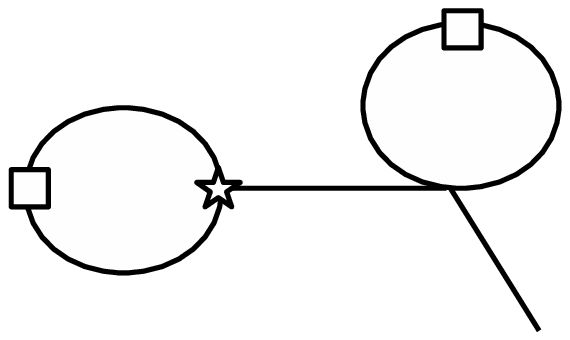,width=6cm}
\caption{Spurious (non 1-PI) contribution to $\G^{(2,0)}[\phi JK_0K_0]$.}
\label{fig.3}
\end{center}
\end{figure}

The last terms in eq.(\ref{appA:amp.1}) 
and in eq.(\ref{appA:amp.2}) subtract the connected
but not 1-PI contributions, shown in Figure ~\ref{fig.2} and 
\ref{fig.3}, that enter in the first term on
the R.H.S. of eqs.~(\ref{appA:amp.1}) and (\ref{appA:amp.2})
respectively.

By using eqs.(\ref{appA:simp.2}),
(\ref{appA:n1}), (\ref{appA:simp.3}) and (\ref{appA:simp.4}) and the fact that
$\G^{(2,0)}[JK_0K_0]$ is zero by $SU(2)$ global symmetry
a straightforward computation shows  that the sum of
eqs.(\ref{appA:ins.1})-(\ref{appA:ins.5}) 
yields
\begin{eqnarray}
&& i \frac{m_D^2}{2} \partial^\mu \frac{\delta \G^{(2,0)}[JJK_0K_0]}{\delta J^\mu_a(x)} 
+ i m_D \frac{\delta \G^{(2,0)}[\phi J K_0 K_0]}{\delta \phi_a(x)} \nonumber \\
&& + i \frac{\delta \G^{(1,0)}[K_0K_0]}{\delta K_0(x)} 
\frac{\delta \G^{(1,0)}[\phi K_0 J]}{\delta \phi_a(x)} 
   + i \frac{\delta \G^{(2,0)}[K_0 K_0 K_0]}{\delta K_0(x)} \frac{\delta \G^{(0)}[\phi J]}{\delta \phi_a(x)} 
   = 0 
\, , \nonumber \\
\label{appA:sum.1}
\end{eqnarray}
i.e. the functional equation for the unsubtracted amplitudes.

\section{Diagrammatic analysis of the functional equation for amplitudes with one counterterm insertion}\label{appB}

We evaluate the insertions in eq.(\ref{sec.3:e7}) and eq.(\ref{sec.3:e8}) in the relevant sector
spanned by two $K_0$ and one $J_\mu$.

The one-loop functional equation for the counterterms (\ref{sec.3:e4}) yields
\begin{eqnarray}
&& m_D \frac{\delta \hG^{(1)}[K_0 J \phi \phi \phi] }{\delta \phi_a(x)} 
+ \frac{\delta \hG^{(1)}[K_0 K_0]}{\delta K_0(x)} \frac{\delta \G^{(0)}[\phi\phi\phi J]}{\delta \phi_a(x)} 
+ \frac{\delta \hG^{(1)}[\phi K_0 J]}{\delta K_0(x)} \frac{\delta \G^{(0)}[\phi\phi K_0]}{\delta \phi_a(x)} 
\nonumber \\
&& 
+ \frac{\delta \G^{(0)}[\phi\phi K_0]}{\delta K_0(x)} \frac{\delta \hG^{(1)}[\phi K_0 J]}{\delta \phi_a(x)}  
+ \frac{\delta \hG^{(1)}[K_0 K_0 \phi \phi]}{\delta K_0(x)} \frac{\delta \G^{(0)}[\phi_a J]}{\delta \phi_a(x)}   + \frac{m_D^2}{2} \partial^\mu \frac{\delta \hG^{(1)}[K_0 JJ\phi\phi]}{\delta J^\mu_a(x)} 
\nonumber \\
&& -2 g \epsilon_{abc} \frac{\delta \hG^{(1)}[K_0 J \phi\phi]}{\delta J^\mu_c(x)} J^\mu_b(x) 
 + g \epsilon_{abc} \frac{\delta \hG^{(1)}[K_0 J \phi\phi]}{\delta \phi_b(x)} \phi_c(x) = 0 \, ,
\label{appB:e1}
\end{eqnarray}
\begin{eqnarray}
&& m_D \frac{\delta \hG^{(1)}[J \phi\phi\phi]}{\delta \phi_a(x)}  
 + \frac{\delta \hG^{(1)}[K_0 J \phi]}{\delta K_0(x)}
                 \frac{\delta \G^{(0)}[\phi\phi]}{\delta \phi_a(x)}
 + \frac{\delta \hG^{(1)}[K_0 \phi\phi]}{\delta K_0(x)}
   \frac{\delta \G^{(0)}[\phi J]}{\delta \phi_a(x)} \nonumber \\
&& + \frac{m_D^2}{2} \partial^\mu \frac{\delta \hG^{(1)}[JJ\phi\phi]}{\delta J^\mu_a(x)} 
   - 2 g \epsilon_{abc} \frac{\delta \hG^{(1)}[J\phi\phi]}{\delta J^\mu_c(x)} J^\mu_b(x) 
   + g \epsilon_{abc} \frac{\delta \hG^{(1)}[J \phi\phi]}{\delta \phi_b(x)} \phi_c(x) = 0 \, .
   \nonumber \\
\label{appB:e2}
\end{eqnarray}

The zero-loop functional equation (\ref{sec.2:e4}) yields
\begin{eqnarray}
&& m_D \frac{\delta \G^{(0)}[J\phi\phi\phi]}{\delta \phi_a(x)}  
- 2 g \epsilon_{abc} \frac{\delta \G^{(0)}[J\phi\phi]}{\delta J^\mu_c(x)} J^\mu_b(x)  \nonumber \\
&& + g \epsilon_{abc} \frac{\delta \G^{(0)}[J\phi\phi]}{\delta \phi_b(x)}\phi_c(x)
   + \frac{\delta \G^{(0)}[K_0 \phi\phi]}{\delta K_0(x)} 
      \frac{\delta \G^{(0)}[J\phi]}{\delta \phi_a(x)} = 0 \, .
\label{appB:e3}
\end{eqnarray}

The insertion of eq.(\ref{appB:e1}) gives 
\begin{eqnarray}
&& - \Big \langle \Big ( m_D \frac{\delta \hG^{(1)}[K_0 J \phi \phi \phi] }{\delta \phi_a(x)} 
+ \frac{\delta \hG^{(1)}[K_0 K_0]}{\delta K_0(x)} \frac{\delta \G^{(0)}[\phi\phi\phi J]}{\delta \phi_a(x)} 
+ \frac{\delta \hG^{(1)}[\phi K_0 J]}{\delta K_0(x)} \frac{\delta \G^{(0)}[\phi\phi K_0]}{\delta \phi_a(x)} 
\nonumber \\
&& 
+ \frac{\delta \G^{(0)}[\phi\phi K_0]}{\delta K_0(x)} \frac{\delta \hG^{(1)}[\phi K_0 J]}{\delta \phi_a(x)}  
+ \frac{\delta \hG^{(1)}[K_0 K_0 \phi \phi]}{\delta K_0(x)} \frac{\delta \G^{(0)}[\phi_a J]}{\delta \phi_a(x)}   + \frac{m_D^2}{2} \partial^\mu \frac{\delta \hG^{(1)}[K_0 JJ\phi\phi]}{\delta J^\mu_a(x)} 
\nonumber \\
&& -2 g \epsilon_{abc} \frac{\delta \hG^{(1)}[K_0 J \phi\phi]}{\delta J^\mu_c(x)} J^\mu_b(x) 
 + g \epsilon_{abc} \frac{\delta \hG^{(1)}[K_0 J \phi\phi]}{\delta \phi_b(x)} \phi_c(x) \Big ) \G^{(0)}[K_0\phi\phi] \Big \rangle = 0 \, .
\label{appB:e4}
\end{eqnarray}

Moreover the insertion in eq.(\ref{appB:e2}) yields
\begin{eqnarray}
&& 
\!\!\!\!\!\!\!\!\!\!\!\!\!\!\!\!
- \frac{i}{2} \Big \langle \Big ( m_D \frac{\delta \hG^{(1)}[J \phi\phi\phi]}{\delta \phi_a(x)}  
 + \frac{\delta \hG^{(1)}[K_0 J \phi]}{\delta K_0(x)}
                 \frac{\delta \G^{(0)}[\phi\phi]}{\delta \phi_a(x)}
 + \frac{\delta \hG^{(1)}[K_0 \phi\phi]}{\delta K_0(x)}
   \frac{\delta \G^{(0)}[\phi J]}{\delta \phi_a(x)} \nonumber \\
&& + \frac{m_D^2}{2} \partial^\mu \frac{\delta \hG^{(1)}[JJ\phi\phi]}{\delta J^\mu_a(x)} 
   - 2 g \epsilon_{abc} \frac{\delta \hG^{(1)}[J\phi\phi]}{\delta J^\mu_c(x)} J^\mu_b(x) 
   \nonumber \\
&& + g \epsilon_{abc} \frac{\delta \hG^{(1)}[J \phi\phi]}{\delta \phi_b(x)} \phi_c(x)
   \Big ) \G^{(0)}[K_0\phi\phi] \G^{(0)}[K_0\phi\phi] \Big \rangle = 0 \, .
\label{appB:e5}
\end{eqnarray}

The insertion in eq.(\ref{appB:e3}) finally gives
\begin{eqnarray}
&& - \Big \langle \Big ( m_D \frac{\delta \G^{(0)}[J\phi\phi\phi]}{\delta \phi_a(x)}  
- 2 g \epsilon_{abc} \frac{\delta \G^{(0)}[J\phi\phi]}{\delta J^\mu_c(x)} J^\mu_b(x)  \nonumber \\
&& + g \epsilon_{abc} \frac{\delta \G^{(0)}[J\phi\phi]}{\delta \phi_b(x)}\phi_c(x)
   + \frac{\delta \G^{(0)}[K_0 \phi\phi]}{\delta K_0(x)} 
      \frac{\delta \G^{(0)}[J\phi]}{\delta \phi_a(x)} \Big ) \hG^{(1)}[K_0 K_0 \phi\phi] \Big \rangle = 0 \,
      \nonumber \\
\label{appB:e6}
\end{eqnarray}
and
\begin{eqnarray}
&& -i \Big \langle \Big ( m_D \frac{\delta \G^{(0)}[J\phi\phi\phi]}{\delta \phi_a(x)}  
- 2 g \epsilon_{abc} \frac{\delta \G^{(0)}[J\phi\phi]}{\delta J^\mu_c(x)} J^\mu_b(x)  \nonumber \\
&& + g \epsilon_{abc} \frac{\delta \G^{(0)}[J\phi\phi]}{\delta \phi_b(x)}\phi_c(x)
   + \frac{\delta \G^{(0)}[K_0 \phi\phi]}{\delta K_0(x)} 
      \frac{\delta \G^{(0)}[J\phi]}{\delta \phi_a(x)} \Big ) \hG^{(1)}[K_0\phi\phi] \G^{(0)}[K_0\phi\phi] \Big \rangle = 0 \, . \nonumber \\
\label{appB:e7}
\end{eqnarray}

The second term in the L.H.S. of eq.(\ref{appB:e4}) yields
\begin{eqnarray}
i \frac{\delta \hG^{(1)}[K_0K_0]}{\delta K_0(x)}
         \frac{\delta \G^{(1,0)}[JK_0\phi]}{\delta \phi_a(x)} 
\label{appB:e8}
\end{eqnarray}
since
\begin{eqnarray}
i \frac{\delta \G^{(1,0)}[JK_0\phi]}{\delta \phi_a(x)} = 
- \Big \langle \frac{\delta \G^{(0)}[J\phi\phi\phi]}{\delta \phi_a(x)} \G^{(0)}[K_0\phi\phi] \Big \rangle \, .
\label{appB:e9}
\end{eqnarray}

The fourth term in the L.H.S. eq.(\ref{appB:e4}) yields
\begin{eqnarray}
i \frac{\delta \hG^{(1)}[JK_0\phi]}{\delta \phi_a(x)} \frac{\delta \G^{(1,0)}[K_0K_0]}{\delta K_0(x)} 
\label{appB:e10}
\end{eqnarray}
since
\begin{eqnarray}
i \frac{\delta  \G^{(1,0)}[K_0K_0]}{\delta K_0(x)}  = - \Big \langle 
\frac{\delta \G^{(0)}[K_0\phi\phi]}{\delta K_0(x)} \G^{(0)}[K_0\phi\phi] \Big \rangle \, .
\label{appB:e11}
\end{eqnarray}

The terms proportional to $\frac{\delta \G^{(0)}[J\phi]}{\delta \phi_a(x)}$ in
eqs.(\ref{appB:e4}), (\ref{appB:e5}), (\ref{appB:e6}) and (\ref{appB:e7}) yield
\begin{eqnarray}
i \frac{\delta \G^{(2,1)}[K_0K_0K_0]}{\delta K_0(x)}  \frac{\delta \G^{(0)}[\phi J]}{\delta \phi_a(x)}
\label{appB:e12}
\end{eqnarray}
as one can see by taking the derivative w.r.t. $K_0(x)$ of
\begin{eqnarray}
i \G^{(2,1)}[K_0K_0K_0] & = & - \Big \langle \hG^{(1)}[K_0K_0\phi\phi] \G^{(0)}[K_0\phi\phi] \Big \rangle
\nonumber \\
                      &   & - \frac{i}{2}  \Big \langle \hG^{(1)}[K_0 \phi\phi] \G^{(0)}[K_0\phi\phi] \G^{(0)}[K_0\phi\phi] \Big \rangle \, .
\label{appB:e12.2}
\end{eqnarray}
The second term in the L.H.S. of eq.(\ref{appB:e5}) contains $\frac{\delta \G^{(0)}[\phi\phi]}{\delta \phi_a(x)} = 
-\square \phi_a(x)$. By making use of eq.(\ref{appA:simp.1})
it can be easily checked by a direct computation that the amplitudes involving the second term in the L.H.S. of eq.(\ref{appB:e5}) cancel out with those involving the third term in eq.(\ref{appB:e4}).

We now evaluate the two-loop amplitudes corrected with the one-loop counterterms  
which enter in the first line of eq.(\ref{sec.3:e1}):
\begin{eqnarray}
i \frac{m_D^2}{2} \partial^\mu \frac{\delta \G^{(2,1)}[JJK_0K_0] }{\delta J^\mu_a}
& = & - \Big \langle \frac{m_D^2}{2} \partial^\mu \frac{\delta \hG^{(1)}[JJK_0\phi\phi]}{\delta J^\mu_a} \G^{(0)}[K_0\phi\phi] \Big \rangle \nonumber \\
                     & & -\frac{i}{2} \Big \langle \frac{m_D^2}{2} \partial^\mu \frac{\delta \hG^{(1)}[JJ\phi\phi]}{\delta J^\mu_a} \G^{(0)}[K_0\phi\phi] \G^{(0)}[K_0\phi\phi] \Big \rangle \nonumber \\
                     & & -i \Big \langle \frac{m_D^2}{2}  \partial^\mu \frac{\delta \hG^{(1)}[JK_0\phi\phi]}{\delta J^\mu_a} \G^{(0)}[J\phi\phi] \G^{(0)}[K_0\phi\phi] \Big \rangle                     
\nonumber \\
  & & -i \Big \langle \hG^{(1)}[JK_0\phi\phi] ~ \frac{m_D^2}{2} \partial^\mu \frac{\delta \G^{(0)}[J\phi\phi]}{\delta J^\mu_a} \G^{(0)}[K_0\phi\phi] \Big \rangle                     
\nonumber \\
                     & & -i \Big \langle \hG^{(1)}[K_0K_0\phi\phi] ~ \frac{m_D^2}{2} \partial^\mu
                                \frac{\delta \G^{(0)}[J\phi\phi]}{\delta J^\mu_a} \G^{(0)}[J\phi\phi] \Big \rangle
\nonumber \\
										 & & +\frac{1}{2} \Big \langle \frac{m_D^2}{2} \partial^\mu 
										                  \frac{\delta \hG^{(1)}[J\phi\phi]}{\delta J^\mu_a}
										                  \G^{(0)}[J\phi\phi] \G^{(0)}[K_0\phi\phi]                                                                          \G^{(0)}[K_0\phi\phi] \Big \rangle \nonumber \\
										 & & +\frac{1}{2} \Big \langle \hG^{(1)}[J\phi\phi] ~
										                  \frac{m_D^2}{2} \partial^\mu 
										                  \frac{\delta \G^{(0)}[J\phi\phi]}{\delta J^\mu_a} \G^{(0)}[K_0\phi\phi]                                             \G^{(0)}[K_0\phi\phi] \Big \rangle \nonumber \\
                     & & + \Big \langle \hG^{(1)}[K_0\phi\phi] ~ \frac{m_D^2}{2} \partial^\mu 
                                       \frac{\delta \G^{(0)}[J\phi\phi]}{\delta J^\mu_a} \G^{(0)}[J\phi\phi] \G^{(0)}[K_0\phi\phi] \Big \rangle \, , \nonumber \\
\label{appB:amp.1}
\end{eqnarray}
\begin{eqnarray}
i m_D \frac{\delta \G^{(2,1)}[\phi J K_0 K_0]}{\delta \phi_a(x)} & = & 
~~ - \Big \langle \G^{(0)}[K_0 \phi\phi] ~  m_D \frac{\delta \hG^{(1)}[K_0 J \phi\phi\phi]}{\delta \phi_a(x)} \Big \rangle
\nonumber \\
& &  -\frac{i}{2} \Big \langle \G^{(0)}[K_0 \phi\phi] \G^{(0)}[K_0 \phi\phi] 
           ~ m_D \frac{\delta \hG^{(1)}[J \phi\phi\phi]}{\delta \phi_a(x)} \Big \rangle \nonumber \\
& &  - \Big \langle m_D \frac{\delta \G^{(0)}[J \phi\phi\phi]}{\delta \phi_a(x)} \hG^{(1)}[K_0 K_0 \phi\phi] \Big \rangle
\nonumber \\
& &  -i \Big \langle m_D \frac{\delta \G^{(0)}[J \phi\phi\phi]}{\delta \phi_a(x)} \G^{(0)}[K_0 \phi\phi] \hG^{(1)}[K_0 \phi\phi] \Big \rangle \, \, . \nonumber \\
\label{appB:amp.1.2}
\end{eqnarray}

In order to proceed further one needs to use eq.(\ref{appA:2}) into eq.(\ref{appB:amp.1}).
This allows to replace the insertion of 
$$\frac{m_D^2}{2} \partial^\mu \frac{\delta \G^{(0)}[J\phi\phi]}{\delta J^\mu_a}$$ 
with that of
$$-g \epsilon_{abc} \frac{\delta \G^{(0)}[\phi\phi]}{\delta \phi_b(x)} \phi_c(x) = 
   g \epsilon_{abc} \square \phi_b(x) \phi_c(x)\, .$$
One can then perform in a straightforward way the relevant contractions with the help of eq.(\ref{appA:simp.1}).
It is also convenient to use the fact that, as a consquence of the one-loop functional equation,
the following identities hold:
\begin{eqnarray}
&& \frac{m_D^2}{2} \partial_\mu \frac{\delta \hG^{(1)}[K_0J\phi\phi]}{\delta J^\mu_a(x)}
+ g \epsilon_{abc} \frac{\delta \hG^{(1)}[K_0\phi\phi]}{\delta \phi_b(x)} \phi_c(x) = 0 \, 
\label{appB:amp.2}
\end{eqnarray}
and
\begin{eqnarray}
\frac{m_D^2}{2} \partial_\mu \frac{\delta \hG^{(1)}[J\phi\phi]}{\delta J^\mu_a(x)} = 0 \, .
\label{appB:amp.3}
\end{eqnarray}
In the above equation we have used the fact that $\hG^{(1)}_{\phi^a \phi^b}=0$ (see
Refs.~\cite{Ferrari:2005ii},\cite{Ferrari:2005va}).
Therefore one obtains
\begin{eqnarray}
i \frac{m_D^2}{2} \partial^\mu \frac{\delta \G^{(2,1)}[JJK_0K_0] }{\delta J^\mu_a(x)}
& = & - \Big \langle \frac{m_D^2}{2} \partial^\mu \frac{\delta \hG^{(1)}[JJK_0\phi\phi]}{\delta J^\mu_a(x)} \G^{(0)}[K_0\phi\phi] \Big \rangle \nonumber \\
                     & & -\frac{i}{2} \Big \langle \frac{m_D^2}{2} \partial^\mu \frac{\delta \hG^{(1)}[JJ\phi\phi]}{\delta J^\mu_a(x)} \G^{(0)}[K_0\phi\phi] \G^{(0)}[K_0\phi\phi] \Big \rangle \nonumber \\
                      & &
                       - \Big \langle g \epsilon_{abc} \frac{\delta \hG^{(1)}[JK_0\phi\phi]}{\delta \phi_b(x)} \phi_c(x) \G^{(0)}[K_0\phi\phi] \Big \rangle                     
\nonumber \\
                     & & 
                      - \Big \langle  ~
                        g \epsilon_{abc} \frac{\delta \G^{(0)}[J\phi\phi]}{\delta \phi_b(x)} \phi_c(x) 
                        \hG^{(1)}[K_0K_0\phi\phi]
                        \Big \rangle
\nonumber \\
										 & & 
                     - \frac{i}{2} \Big \langle 
                       g \epsilon_{abc} \frac{\delta \hG^{(1)}[J\phi\phi]}{\delta \phi_b(x)} \phi_c(x)
                     \G^{(0)}[K_0\phi\phi] \G^{(0)}[K_0\phi\phi] \Big \rangle \nonumber \\
                     & & 
                     -i \Big \langle  g \epsilon_{abc} \frac{\delta \G^{(0)}[J\phi\phi]}{\delta \phi_b(x)} \phi_c(x)
                         \hG^{(1)}[K_0\phi\phi] \G^{(0)}[K_0\phi\phi] \Big \rangle \, . \nonumber \\
\label{appB:amp.1.simp}
\end{eqnarray}
By taking into account eqs.(\ref{appB:e8}),(\ref{appB:e10}),(\ref{appB:e12}),
(\ref{appB:amp.1.2}) and (\ref{appB:amp.1.simp}) and the fact that $\G^{(2,1)}[K_0K_0J]$ is zero
by $SU(2)$ global symmetry the sum of eqs.(\ref{appB:e4}),(\ref{appB:e5}),(\ref{appB:e6}) and
(\ref{appB:e7}) yields finally
\begin{eqnarray}
&& 
i \Big ( \frac{m_D^2}{2} \partial^\mu \G^{(2,1)}_{J^\mu_a}[JJ K_0 K_0] + m_D \G^{(2,1)}_{\phi_a}[K_0 K_0 J \phi]
\nonumber \\
&& 
+ \G^{(2,1)}_{K_0} [K_0 K_0 K_0] \G^{(0)}_{\phi_a} [\phi J]
 + \G^{(1,0)}_{K_0}[K_0 K_0] \hat \G^{(1)}_{\phi_a}[\phi K_0 J] \nonumber \\
&& + \hG^{(1)}_{K_0} [K_0K_0] \G^{(1,0)}_{\phi_a} [J K_0 \phi] \Big ) = 0 \, .
\label{appB:final}
\end{eqnarray}

\end{document}